\documentclass[aps]{article}
\usepackage{epsfig}
\newcommand{\be}{\begin{eqnarray}}
\newcommand{\ee}{\end{eqnarray}}
\newcommand{\la}{\langle}
\newcommand{\ra}{\rangle}
\newcommand{\tra}{{\rm \hat Tr}} 
\newcommand{\del}{\partial}
\begin{document}
\vskip 1cm
\centerline{\Large \bf Complex Langevin Equation and the Many-Fermion Problem}
\vskip 1cm
\centerline{\large Chris Adami and Steven E. Koonin}
\vskip 0.5cm
\centerline{\it W. K. Kellogg Radiation Laboratory, California  
Institute of Technology}
\centerline{\it Pasadena, California 91125 USA}
\vskip 2cm
\begin{abstract}
We study the utility of a complex Langevin (CL) equation as an alternative
for the Monte Carlo (MC) procedure in the evaluation of expectation values
occurring in fermionic many-body problems. We find that  
a CL approach is natural in cases where non-positive definite
probability measures occur, and remains accurate even when the
corresponding MC calculation develops a severe ``sign problem''. While
the convergence of CL averages cannot be guaranteed in principle,
we show how convergent results can be obtained in three examples
ranging from simple one-dimensional integrals over quantum mechanical
models to a schematic shell model path integral.
\end{abstract}
\vskip 6cm
\noindent Corresponding author:
\vskip 0.15in
Chris Adami

E-mail: adami@caltech.edu

Phone (+) 626 395-4256

Fax (+) 626 564-8708
\vskip 1cm

\newpage
\section{Introduction}
There has been recent significant progress in the large-scale
numerical computation of nuclear properties in the shell model, using
the Auxiliary Field Path-Integral (AFPI) Monte Carlo method
\cite{LJKO}. Due to the benign scaling of the computational effort
with the single-particle basis, properties of large nuclei can now be
calculated that are out of the reach of conventional diagonalization
methods \cite{KDLa,KDLb}.

Quite generally, AFPI treatments of interacting fermion systems with
the Monte Carlo (MC) method are difficult for certain realistic
Hamiltonians, and for odd-particle configurations\footnote{See for
example~\cite{SUZ}}. This difficulty, also known as the ``sign
problem'', is the prime impediment to large scale computational
efforts both in nuclear and condensed matter physics calculations.
Briefly, repulsive interactions and/or odd-particle configurations can
lead to probability distributions (integration measures for the
auxiliary fields) that are negative, or even complex. As the MC
update algorithm relies on a positive-definite measure, the sign of
the distribution is made part of the observable being calculated.
Under circumstances where the average sign of the distribution is
small, the expectation value is the ratio of two very small numbers,
that converges only asymptotically.

A number of alternatives to or extensions of the MC method have been
proposed over the years, among them hybrid methods combining the MC
method with replication mechanisms for importance sampling~\cite{DEU},
random-walk branching \cite{ZCG}, and diagonalization over optimal
bases using variational techniques~\cite{HMO}. Here, we investigate
the possibility of replacing the MC method altogether with one based
on the complex Langevin (CL) equation, at least in those cases where
the sign problem is prominent. The CL equation has received
considerable attention in connection with lattice gauge theory
calculations, where either static charges or a non-zero chemical
potential give rise to complex actions~\cite{CLREF}. It has been
abandoned mostly because of the perception that complex Langevin
averages ought not to be trusted, due to the possibility that they can
be non-stationary. Here, we show that the problem of non-stationarity
can often be avoided, and does not play a role in a large class of
Hamiltonians which give rise to Langevin equations with fixed points
in the complex plane.

By way of introduction and to establish notation, we briefly review 
the AFPI Monte Carlo method in the next section. Section 3 discusses
the application of the CL method to simple one-dimensional integrals
abstracted from those occurring in fermionic many-body systems,
while Section 4 expands this to a simple toy Hamiltonian with 
characteristics reminiscent of shell models. In Section 5 we apply the
method to the Lipkin model (the MC treatment of which is similar in character
to the full shell model), and close with conclusions, in Section 6.

\section{Auxiliary Field Path-Integral Monte Carlo}
\setcounter{equation}{0}
In the AFPI method, the significant savings in computational effort
are obtained through a Hubbard-Stratonovich (HS) transformation~\cite{HS} (see
below) of the imaginary-time evolution operator
\be
\hat U = \exp(-\beta \hat H) \label{evop}\;.
\ee
The thermal expectation value of an operator $\hat O$ is given by
\be
\la\hat O\ra_\beta = 
Z^{-1}\ \tra \left[\hat O\ \exp{(-\beta \hat H)}\right]\;,
\ee
where 
\be
Z = \tra \exp(-\beta \hat H)
\ee
is the partition function, $\tra$ is the many-body trace, and $\beta$
is the inverse temperature. Ground-state (zero-temperature) properties are 
obtained in the limit $\beta\rightarrow\infty$. 

For a many-body operator $\hat H$ in the quadratic form
\be
\hat H =\sum_\alpha\epsilon_\alpha\hat{\cal O}_\alpha + \frac12 \sum_\alpha
V_\alpha\hat{\cal O}^2_\alpha \label{ham}
\ee
containing one-body operators $\hat{\cal O}_\alpha$ and two-body operators
$\hat{\cal O}_\alpha^2$, a HS transformation of the imaginary-time evolution 
operator (\ref{evop}) leads to
\be
\exp(-\beta \hat H)\approx\int_{-\infty}^{\infty}\prod_{\alpha,n}
d\sigma_{\alpha,n}\left(\frac{\Delta\beta|V_\alpha|}{2\pi}\right)^{1/2}
G(\sigma)\, \prod_n\, \exp(-\Delta\beta\,\hat h(\vec\sigma_n))\;,
\ee
by splitting $\beta$ into $N_t$ time slices such that $\Delta\beta=\beta/N_t$.

Here, $\vec \sigma_n$ denotes a set of auxiliary fields (one for each 
two-body operator appearing in (\ref{ham})) at time-slice $n$, and 
$\sigma$ stands for the totality of fields at all time-slices. Furthermore, 
$G(\sigma)$ is the gaussian weight factor
\be
G(\sigma) = \exp(-\sum_{\alpha,n}\frac12\, \Delta\beta\, |V_\alpha|\, 
\sigma_{\alpha n}^2 )\;,
\ee
and $\hat h(\vec\sigma_n)$ is the {\em one-body} Hamiltonian
\be
\hat h(\vec\sigma_n) = \sum_\alpha\left(\epsilon_\alpha
+s_\alpha V_\alpha\sigma_{\alpha n}\right)\hat {\cal O}_\alpha \label{obham}
\ee
where $s_\alpha=\pm1\ \ (=\pm {\rm i})$ if $V_\alpha<0\ \ (>0)$. Thus, the HS 
transformation has the effect of replacing the quadratic dependence on 
$\hat {\cal O}_\alpha$ in (\ref{ham}) with a linear one, at the expense of an
integral over auxiliary fields.

Let $\hat U_\sigma$ denote the {\em one-body} evolution operator
\be
\hat U_\sigma = \exp(-\Delta\beta\,\hat h(\vec\sigma_n)) 
\ee
and $F(\sigma)$ its trace:
\be F(\sigma) = \tra\, (\hat U_\sigma)\;.
\ee
Expectation values can then be written using the above path-integral
decomposition of the partition function:
\be
\la \hat O\ra_\beta = 
\frac{\int{\cal D}[\sigma]\, G(\sigma)\,  \tra\, ( \hat O\, \hat U_\sigma)}
{\int {\cal D}[\sigma]\, G(\sigma)\, F(\sigma)} \;.\label{expval}
\ee
An effective action $S_\sigma$ can be defined such that
(\ref{expval}) appears as a simple expectation value:
\be
\la \hat O\ra_\beta = 
\frac{\int{\cal D}[\sigma]\, e^{-S_\sigma}\, \la \hat O\ra_\sigma}
{\int {\cal D}[\sigma]\ e^{-S_\sigma} } 
\ee
with 
\be
S_\sigma = \sum_{\alpha,n}\frac12 \Delta\beta|V_\alpha|\sigma_{\alpha,n}^2
-\log\,F(\sigma)\;,
\ee
where 
$\la \hat O\ra_\sigma = \tra\, ( \hat O\, \hat U_\sigma)/F(\sigma)\;$.

Further, since $\exp(-S_\sigma)$ is not positive definite, we define the 
sign-function
\be
\Phi_\sigma = \frac{F(\sigma)}{|F(\sigma)|}\;.
\ee
Then, writing $e^{-S_\sigma} = \Phi(\sigma)\, e^{-\tilde S_\sigma}$
with a {\em real} $\tilde S_\sigma$, the expectation value (\ref{expval})
appears as a {\em ratio} of expectation values
\be 
\la \hat O\ra_\beta = 
\frac{\int{\cal D}[\sigma]\ e^{-\tilde S_\sigma}\, 
\Phi(\sigma)\,\la\hat O \ra_\sigma}
{\int {\cal D}[\sigma]\, e^{-\tilde S_\sigma}\,\Phi(\sigma)}
= \frac {\ll\Phi_\sigma \la\hat O\ra_\sigma \gg}{\ll \Phi_\sigma\gg}  
\ee
each calculated  with a positive definite probability distribution 
$e^{-\tilde S_\sigma}$. The Monte Carlo average over samples is
denoted as $\ll...\gg$. If the function $F(\sigma)$ is negative on
a substantial part of the $\sigma$-field manifold, the expectation
values $\ll \Phi_\sigma \la\hat O\ra_\sigma \gg$ and 
\mbox{$\ll \Phi_\sigma\gg$}
 can each become very small, and the ratio only converges 
asymptotically. This is the essence of the sign problem in
quantum Monte Carlo calculations. Below, we construct expectation
values susceptible to the sign problem, in order to gain insight into 
the circumstances in which a CL equation approach can be applied successfully.

\section{Complex Langevin Equation and Simple Integrals}
\setcounter{equation}{0}
The HS transformation is nothing but the Gaussian integral identity
\be
e^{-\frac12z^2} = \frac1{\sqrt{2\pi}}\int_{-\infty}^{\infty}
d\sigma\,e^{-\frac12\sigma^2}e^{\pm {\rm i}\sigma z}\;.  \label{hs}
\ee
Note that this case corresponds to a repulsive interaction [$V>0$ 
in (\ref{obham})]. As (\ref{hs}) has no imaginary part, this reduces to
\be
e^{-\frac12z^2} = \frac1{\sqrt{2\pi}}\int_{-\infty}^{\infty}
d\sigma\,e^{-\frac12\sigma^2}\cos(z\sigma)\;.
\ee
This is precisely the form appearing in the denominator of (\ref{expval}),
but in one dimension, with $\cos(z\sigma)$ playing the role of $F(\sigma)$.

Generalizing this, we would like to examine expectation values
\be
\la\sigma^2\ra_N = \frac{\int d\sigma\  \sigma^2 
e^{-\frac12\sigma^2}\,[\cos(\sigma z)]^N}
{\int d\sigma\  e^{-\frac12\sigma^2}\,[\cos(\sigma z)]^N}\label{square}
\ee
as a function of the real number $z$ and the ``particle number'' $N$,
using MC evaluation and the complex Langevin equation approach.

Clearly, the MC procedure will suffer from the sign-problem only for 
odd $N$. Fig.~1 shows a straightforward MC evaluation of this integral 
for $N=1$, where the inset shows the development of the 
average sign $\Phi(\sigma)=\cos(z\sigma)/|\cos(z\sigma)|$. As expected, 
the accuracy of the MC estimate deteriorates as $\Phi(\sigma)\rightarrow0$.
\begin{figure}[tb]
\center
 \epsfig{file={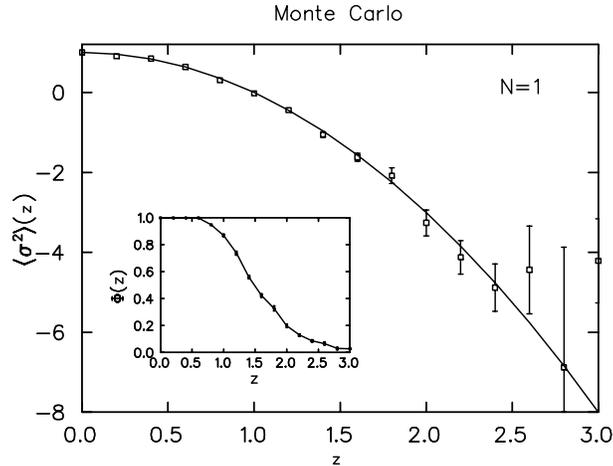}, width = .4\columnwidth, angle=90}
\caption{Monte Carlo average of integral $\la \sigma^2\ra_{N=1}$
(\ref{square}) (squares), the exact solution (solid line), and the
sign of the calculation (insert)}
\end{figure}

As an alternative to the MC procedure, consider the Langevin equation. 
For systems with real actions, expectation values such as
\be
\la O\ra = \frac1Z\int d\sigma\ O(\sigma)\, e^{-S(\sigma)} \label{laneval}
\ee 
with the partition function
\be
Z = \int d\sigma\ e^{-S(\sigma)}
\ee
can be calculated by creating a stochastic process using the Langevin equation,
with an equilibrium distribution $P_0(\sigma) = Z^{-1}\exp(-S(\sigma))$.

The Langevin equation is given by
\be
\frac{d\sigma(t)}{dt}=-\frac12\ \frac{\del S}{\del\sigma}+\eta(t)\;,
\label{rlan} 
\ee
where $t$ is a fictitious time
and $\eta(t)$ is stochastic noise with zero mean and unit variance:
\be
\la\eta(t)\eta(t^\prime)\ra = \delta(t-t^\prime)\;.
\ee
To each Langevin equation corresponds a Fokker-Planck equation for the 
probability density $P(\sigma,t)$
\be
\frac{\del P(\sigma,t)}{\del t} = H_{FP}\ P(\sigma,t)  \label{fp}
\ee
with a Hermitian Fokker-Planck Hamiltonian
\be 
H_{FP} = \frac12\frac\del{\del\sigma}\,
\left(\frac\del{\del\sigma}+ \frac{\del S}{\del\sigma}\right) \label{fpham}\;.
\ee
For solutions with exponential time dependence,
\be
P(\sigma,t) = e^{-E\,t}P_E(\sigma)\;,
\ee
Equation (\ref{fp}) reverts to an eigenvalue equation. For 
$t\rightarrow\infty$, $P(\sigma,t)\rightarrow P_0(\sigma)$, the solution
for the lowest eigenvalue $E=0$. Accordingly, for $t\rightarrow\infty$, if
$\sigma_0(t)$ is the solution to the Langevin equation (\ref{rlan}),
\be
O(\sigma_0(t))\rightarrow \la O \ra\ \ \ (t\rightarrow\infty)\;
\ee
with $\la O\ra$ given by (\ref{laneval}). Finally, ergodicity assures
that $\la O\ra$ is also obtained by averaging over the path $\sigma_0(t)$:
\be
\la O \ra =\ \ll O(\sigma_0)\gg\ = 
\lim_{T\rightarrow\infty}\frac1T\int_0^TO(\sigma_0(t))\;.
\ee

In principle, nothing prevents us from using the Langevin equation in the 
case where the action is complex:
\be
S(\sigma) = S_R(\sigma)+{\rm i}\ S_I(\sigma)\;.
\ee
Then, we obtain two equations, for the real and the imaginary part of $\sigma$:
\be
\frac{\del\sigma_R(t)}{\del t} & = & 
- \frac12\ {\rm Re}\left(\frac{\del S}{\del\sigma}\right)+\eta(t)\label{clr}\\
\frac{\del\sigma_I(t)}{\del t} & = & \label{cli}
- \frac12\ {\rm Im}\left(\frac{\del S}{\del\sigma}\right)\;. 
\ee
However, the Fokker-Planck Hamiltonian loses its hermeticity, and
the eigenvalues can acquire imaginary parts. As a consequence, the 
probability distribution $P(\sigma_R,\sigma_I,t)$ need not 
converge anymore, nor does the expectation value $\la O\ra$.  
While the lowest eigenvalue is still $E_0=0$, the $E_n$ with $n>0$ are in
general complex, and the asymptotic condition $P(\sigma)\rightarrow P_0$
is violated whenever there are any $E_n$ with ${\rm Re}\ E_n>0\ \ (n>0)$ 
\cite{KLA}. As a rule of thumb then, expectation values obtained via
the CL equation should only be trusted if the ensemble averages become 
time-independent~\cite{LEE}. 

For the numerical solution of equations (\ref{clr},\ref{cli}) we use 
the two-step algorithm of Greenside and Helfand \cite{GH}.
Defining the complex gradient as
\be
\nabla S(t)=\frac{\del
  S}{\del\sigma}\left[\sigma_R(t),\sigma_I(t)\right]\;,
\ee
the stochastic differential equation is discretized via
\be
\sigma_R(t_{1/2})&=& \sigma_R(t_0)-\Delta t\ 
{\rm Re}\left[\nabla S(t_0)\right]
+ \sqrt{\Delta t}\,\eta(t_0) \\
\sigma_R(t_1)&=& \sigma_R(t_0)-\frac12\, \Delta t\, {\rm Re}\left[
\nabla S(t_0)+\nabla S(t_{1/2})\right]+
\sqrt{\Delta t}\,\eta(t_0)
\ee
and analogously for the imaginary part\footnote{For this particular
  choice of variables, the corresponding equation for the imaginary
  part does not include a noise term. For other choices, see, e.g.,
  \cite{okamoto89}.}.

For the expectation value (\ref{square}), the effective action is
\be
S_N = \frac12\, \sigma^2 - N\, \log[\cos(\sigma z)] \label{effact}
\ee
while the associated Langevin equation reads
\be
\dot \sigma = -\frac12\,(\sigma+N\, z \,\tan(z\sigma)) +\eta\;.  \label{simpeq}
\ee
\begin{figure}[bt!]
\center
 \epsfig{file={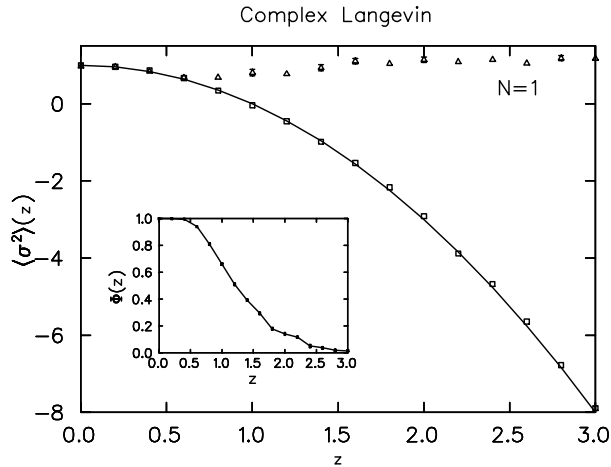}, width = .4\columnwidth, angle=90}
\caption{Langevin average of  $\la \sigma^2\ra_{N=1}$. Squares: 
extended action, triangles: original action, solid line: exact
solution. }
\end{figure}
It is immediately apparent that the fixed points of this equation
$(\dot \sigma=0$) lie on the real axis. On the other hand it also
clear that, since for example $\la\sigma^2\rangle_{N=1}=1-z^2$, the
correct solution $\sigma_0(t)$ needs to spend a considerable amount of
time away from the real line, at least for odd $N$. Analytically, this
must happen because (for odd $N$) there is a delta-function--like
drift term that has been ignored in (\ref{simpeq}) which is due to the
imaginary part of $S_N$. The additional drift term is
\be
\delta\nabla S_N = \pm {\rm i}\pi\delta(\cos(z\sigma))
\ee
but cannot be adequately modeled numerically. As a consequence, the solution 
spends most of its time on the real line between the first turning points,
and the resulting average is inaccurate, as has been noted previously 
\cite{HAY1}. The solution to this dilemma is also not new. Going back to
(\ref{hs}) and $N=1$ we see that the action can also be written as
\be 
S_1 = \frac12\, \sigma^2\, \pm\ {\rm i}z\sigma \label{extact}\;,
\ee
in which case the fixed point is away from the real line in the complex
plane: $\sigma_I = \pm z$. Since then $P(\sigma_R,\sigma_I)\rightarrow
\exp(-\frac12\,\sigma_R^2)\ \delta(\sigma_I\mp z)$, the complex average 
reduces to
\be
\la \sigma^2\ra_{N=1} = \frac{\int_{-\infty}^{\infty}
\sigma^2 P(\sigma_R,\sigma_I)\,d\sigma_R\,d\sigma_I}{\int_{-\infty}^{\infty}
P(\sigma_R,\sigma_I)\,d\sigma_R\,d\sigma_I}
=\frac{\int_{-\infty\mp iz}^{\infty\pm iz}d\sigma\, \sigma^2 
\exp(-\frac12\,\sigma^2)}
{\int_{-\infty\mp iz}^{\infty\pm iz}d\sigma\,\exp(-\frac12\, \sigma^2)}\;.
\ee     
The integral over the complex path removed from the real line by $\pm
{\rm i}z$
equals the one over the real line if the observable has no poles in the 
enclosed area.

In Fig.~2 we show the result of a CL evaluation of
$\la\sigma^2\ra_{N=1}$ using the ``extended'' action (\ref{extact}) (squares)
and the original action (\ref{effact}) (triangles). The solid line
represents the exact result. As expected,
a complex Langevin simulation  with fixed points on the real line does
not converge, while the calculation with the extended action is robust even
in the regime where the MC (Fig.~1) fails. 

\begin{figure}[tb]
\center
 \epsfig{file={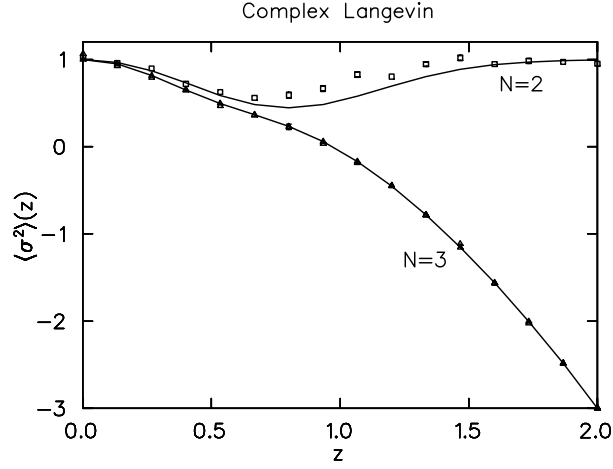}, width = .4\columnwidth, angle=90}
\caption{Complex Langevin average of  $\la \sigma^2\ra_{N}$
for $N=2$ and $N=3$.
}
\end{figure}
Can this procedure
be extended to arbitrary $N$? A canonical extension of the method exists, 
where $\cos^N(\sigma\,z)$ is decomposed into single powers of cosines at 
multiples of $\sigma z$. Subsequently, the cosines are replaced by 
exponentials. However, this procedure results in a shift of the fixed point
away from the real line {\em only} for odd $N$. Fig.~3 shows results for the cases
$N=2$ and $N=3$, with exact results 
\be
\la\sigma^2\ra_{N=2}&=&1-\frac{4\,z^2}{1+e^{2\,z^2}}\;,\\
\la\sigma^2\ra_{N=3}&=&1-z^2\frac{1+3e^{-4\,z^2}}{1+\frac13e^{-4\,z^2}}\;.
\ee
For even $N$ the complex Langevin equation does not converge at all $z$
due to the problems described. We do not need to worry about this, however, 
as the MC procedure is very accurate there. 

\section{Quantum Mechanical Toy Model}
\setcounter{equation}{0} 
In this section we show how shifting the fixed points in a CL
evaluation of integrals can be used in a quantum mechanical model
which describes a single shell of angular momentum $j$ with variable
filling. This toy model is defined by the Hamiltonian
\be
\hat H = -\frac12\,\epsilon\, \hat N^2 + \frac12\, V\, \hat J_z^2\;.
\label{hamilton}
\ee
where $\hat N$ is the number operator and $\hat J_z$ the third
component of the angular momentum.  A HS-transformation on the
imaginary-time evolution operator yields
\be
\tra_N\left(e^{-\beta\, \hat H}\right)\propto \int d\sigma_0\,d\sigma_1\, 
e^{-\frac12\,\left(\sigma_0^2 +\sigma_1^2\right)}\,
\tra_N\left(e^{\sqrt{\beta\epsilon}\,\hat N\,\sigma_0 - {\rm i}\sqrt{\beta V}\,\hat J_z\,\sigma_1}\right)
\ee
where $\tra_N$ is the many-body trace for fixed particle number $N$ and
angular momentum $j$
\be
\tra_N(\hat O)= \sum_{\stackrel{m_1+...+m_N=-j}{m_i\,\neq \,m_j}}^j
\la m_1...m_N |\hat O|m_1...m_N\ra \label{mbtrace}\;.
\ee
Since $\hat N$ and $\hat J_z$ commute, the evolution operator does not need
to be decomposed into time-slices. Also, the contribution from the number 
operator is a constant factor
\be
\tra_N\left(e^{-\beta\,\hat H}\right) = \exp(\sqrt{N\beta\,\epsilon})\,
\tra_N(U_{\sigma_1})
\ee 
that drops out of the ratios, and will thus be ignored in the following.
Above, we defined the one-body evolution operator
\be
U_{\sigma_1} = e^{-{\rm i}\sqrt{\beta\,V}\,\hat J_z\,\sigma_1}\;.
\ee
Defining $\phi = \sqrt{\beta\,V}\sigma_1$ and as before
\be
F_N(\phi)= \tra_N(U_{\sigma_1})\;, \label{tru}
\ee
we can write the $N$-particle traces in terms of the one-particle trace
\be
F_1(\phi) & = & F(\phi) = \frac{\sin(j+\frac12)\phi}{\sin(\phi/2)}\\
F_2(\phi) & = & F^2(\phi)-F(2\phi)\\
F_3(\phi) & = & F^3(\phi)-3\,F(2\phi)\,F(\phi)+2F(3\phi)
\ee
and so on. Note that unlike in the previous section, the even-$N$ trace
is not positive-definite, while still being mostly positive. 

We shall focus on the expectation value\footnote{Note that we
now write $\sigma$ instead of $\sigma_1$ for simplicity.}
\be
\la J_z^2\ra_N = \frac{\int\,d\sigma\,e^{-\frac12\,\sigma^2}\,
\tra_N\left(\hat J_z^2\,U_\sigma\right)}{\int\,d\sigma\,e^{-\frac12\,\sigma^2}
F_N(\sigma)} \label{toyjsq}
\ee
which we rewrite in terms of an effective action as follows
\be
\la \hat J_z^2\ra = \frac{\int\,d\sigma\,e^{-\frac12\,\sigma^2}\,
\la \hat J_z^2\ra_\sigma{\,e^{-S_N(\sigma)}}}
{\int\,d\sigma\,e^{-\frac12\,\sigma^2}\,e^{-S_N(\sigma)}}\;,
\ee
where
\be
\la \hat J_z^2\ra_\sigma = \frac{\tra_N(\hat J_z^2\,U_\beta)}{F_N(\sigma)}
\ee
and
\be 
S_N = \frac12\,\sigma^2\,-\,\log(F_N(\sigma))\;.
\ee
For $N=1$ and arbitrary $j$, we find for the observable 
($\phi = \sqrt{\beta\,V}\sigma$)
\be
\la\hat J_z^2\ra_\sigma = j\,(j+1)\,+\,\cot\frac\phi2
\left((j+\frac12)\,\cot(j+\frac12)\phi\,-\frac12\,\cot\frac\phi2\right)
\label{toyobs}
\ee
while the Langevin equation is 
\be
\dot\sigma = -\frac12\,\sigma\, +\, \frac12\sqrt{\beta\,V}
\left((j+\frac12)\cot(j+\frac12)\phi
-\frac12\,\cot\frac\phi2\right)\,+\eta\;. \label{toylaneq}
\ee

From the oscillatory nature of $F(\phi)$ we expect that the MC
procedure will become imprecise at large $\beta$. In Fig.~4 (left
panel) we show the result of a MC calculation of $\la\hat
J_z^2\ra_\beta$ for a $j=5/2$-shell and particle numbers $N=1,N=2$,
and $N=3$. Note that $N=3$ corresponds to half-filling, so that
higher $N$'s can be described in terms of ``hole''-numbers, and revert
to the cases displayed. For this simple case, the sign (not shown) does not
deteriorate too much before the ground state has been reached
($\beta$ large), and the calculation is consequently reliable. Let us
test nevertheless how a complex Langevin approach fares.

\begin{figure}[tb]
\center
 \epsfig{file={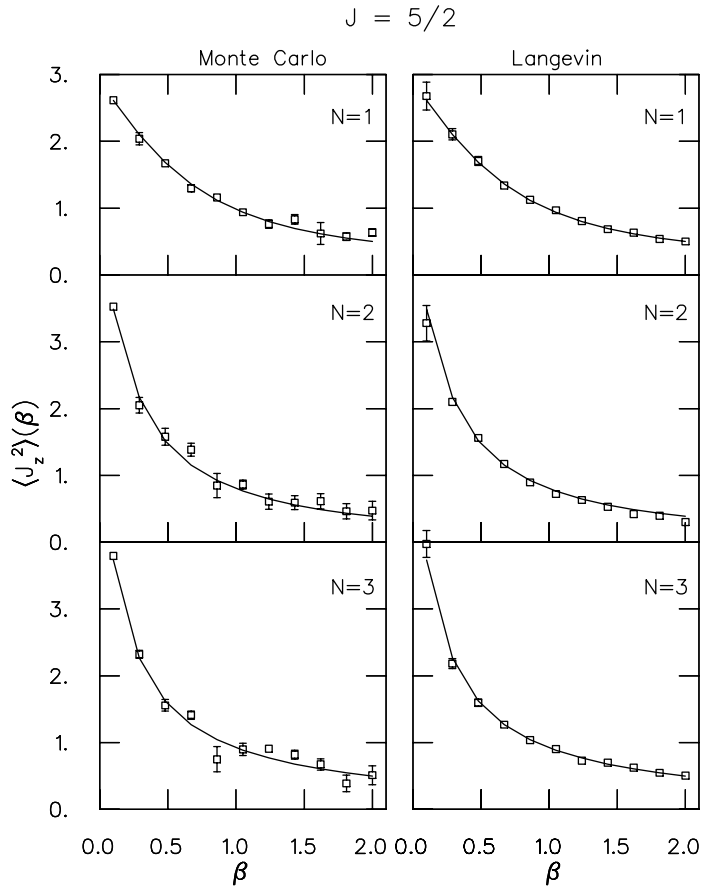}, width = .6\columnwidth, angle=0}
\caption{Average $\la J_z^2\ra$ for a
$j=5/2$ shell with $N=1,2,3$ obtained with a Monte Carlo approach (left panel)
and the complex Langevin equation (right panel). For both cases we
took 10 samples of 10,000 updates each.
}
\end{figure}

The CL approach based on the Langevin equation (\ref{toylaneq}) with
the observable (\ref{toyobs}) suffers from the same problems that we
noted with the simple integral: the fixed points are real and the
results are consequently unreliable. Again, the remedy is to shift the
fixed points such that the effective path of integration lies in the
complex plane. However, here we encounter another difficulty (which
is also common in more refined shell-model calculations): the
expectation value (\ref{toyobs}) has poles (at the zeros of
$F(\phi)$). Consequently, the expectation value calculated for a
shifted path will equal the real-path result plus the sum over the
poles on the real-line, of which there are infinitely many. For this
simple toy Hamiltonian, this can be shown to hold true exactly by
calculating the residues. In
more realistic models, however, the sum over the poles is not readily
available.  Fortunately, for most Hamiltonians the observable
$\tra_N\,(O\,U_\sigma)$ can be obtained by calculating moments of
$\tra_N\,U_\sigma$. Here, for example

\be
\tra_N\,(\hat J_z^2\, U_\sigma)=
-\frac1{\beta\,V}\frac{\del^2}{\del\,\sigma^2}\,\tra_N\,U_\sigma
\ee
and integrating by parts twice yields
\be
\la \hat J_z^2\ra = \frac{\int\,d\sigma\,e^{-\frac12\,\sigma^2}\,
\la \hat J_z^2\ra_\sigma{\,e^{-S_N(\sigma)}}}
{\int\,d\sigma\,e^{-\frac12\,\sigma^2}\,e^{-S_N(\sigma)}}=
\frac1{\beta V}\left(1 -  \frac{\int\,d\sigma\,e^{-\frac12\,\sigma^2}\,
\sigma^2\,{\,e^{-S_N(\sigma)}}}
{\int\,d\sigma\,e^{-\frac12\,\sigma^2}\,e^{-S_N(\sigma)}}\right)
= \frac1{\beta V}(1- \la\sigma^2\ra)\;.
\ee
With this observable, the action can now be extended into the complex
plane.  This is achieved as in the simple integrals treated
previously, by writing the trace in terms of cosines, and replacing
$\cos(\phi)\rightarrow \exp(i\phi)$.  Quantum mechanically, this
amounts to retaining only those terms
in the trace (\ref{mbtrace}) for which the sum of the magnetic quantum
numbers $m_1+m_2+...m_N$ (with $m_i\neq m_j$) is either non-negative
or non-positive. In this manner, we break time-reversal invariance by
hand, since we know that it will be taken care of by the symmetry of
the integral. Results of the CL calculation are shown in the right
panel of Fig.~4.
The CL averages are stable and
accurate at large $\beta$ while accuracy deteriorates only for small
$\beta$ where large cancellations must occur. Of course, a partial
integration can also be performed for the Monte Carlo integral, which
improves performance markedly because of better sampling. In that case,
the Monte Carlo results are comparable to those obtained with the
Langevin equation. 

This changes when the action is forced to acquire substantial complex
pieces by ``cranking'' the Hamiltonian (\ref{hamilton}), which corresponds
to a toy nucleus with a single shell undergoing a collective rotation.
Ignoring the term involving the number operator and again setting
$V_2=1$, the Hamiltonian becomes
\be
\hat H_c = \hat H\,-\,\omega\,\hat J_z  \label{crham}\;,
\ee
which breaks time-reversal invariance explicitly.  Cranking is
notoriously difficult for quantum MC calculations because the
sign-problem is exacerbated in these cases. In fact, a straightforward
MC is hopeless because the sign drops very quickly with increasing
$\omega$. In Fig. 5 we show $\la J_z^2\ra$ as well as the sign for two
particles in a $j=5/2$-shell for $\omega<1.5$, beyond which the MC
calculation becomes useless.  Convergence improves markedly if the
observable is simplified by partial integration (Fig. 6), but of
course the sign is still the same and the MC approach fails.

\begin{figure}[!tbp]
\center
 \epsfig{file={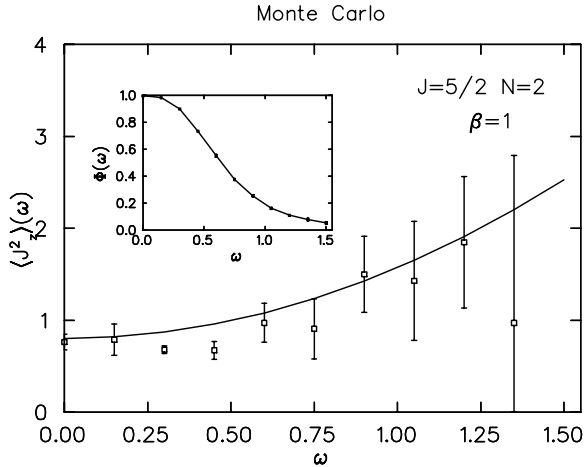}, width = .4\columnwidth, angle=90}
\caption{Straightforward Monte Carlo average of $\la
J_z^2\ra$ for two particles in a cranked $j=5/2$ shell at $\beta=1$, and the
sign (inset). 
}
\end{figure}
\begin{figure}[!tbp]
\center
 \epsfig{file={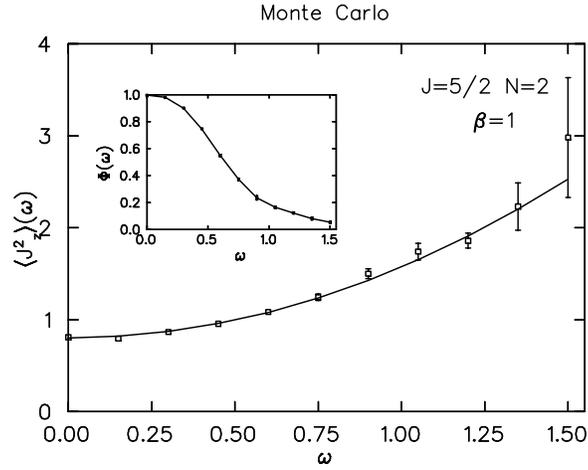}, width = .4\columnwidth, angle=90}
\caption{
Monte Carlo with $\la
J_z^2\ra$ from partial integration, as in Fig. 5.
}
\end{figure}

Let us calculate this observable
with the CL approach. In Fig.~7 we plot the expectation value
(\ref{toyjsq}) as a function of cranking frequency for a $j=5/2$-shell
with $N=1,2,3$ and $\beta = 1.0$. While the sign in the MC calculation 
essentially disappears for $\omega>3$, the accuracy of the
CL calculation is maintained even as the average sign is small.
\begin{figure}[!tbp]
\center
 \epsfig{file={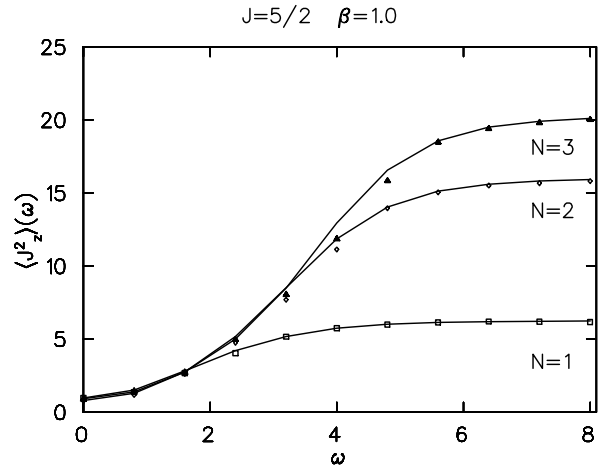}, width = .4\columnwidth, angle=90}
\caption{
Complex Langevin average of $\la J_z^2\ra$ for a
cranked $j=5/2$-shell, at $\beta=1.0$. }
\end{figure}

\section{Lipkin Model}
\setcounter{equation}{0}
To test the CL equation approach in a more realistic
situation for a  system that can be exactly diagonalized, we use 
the Lipkin model \cite{LIP}. The Lipkin model is a 
non-trivial schematic shell model capable of describing collective effects 
in nuclei. It describes $N$ distinguishable particles labeled 
$1,2,...,N$, each of which can occupy one of two orbitals with
energies $(\pm1/2)$ (up or down). The total number of states therefore
is $2^N$. 

The Hamiltonian has a one-body term and two two-body terms,
and suffers from the sign problem as we shall see below. If written in
terms of quasi-spin operators $\hat J$, it becomes
\be
\hat H = \hat J_z -\frac12\, V\left(\hat J_+^2 + \hat J_-^2\right) = 
\hat J_z - V\left( \hat J_x^2 - \hat J_y^2\right)\;.
\ee 
The eigenstates can be labeled by the total quasi-spin $j$, and
classified into non-degenerate multiplets of $2j+1$ states each (see,
e,g, \cite{koo90}), from spin $j=N/2$ down to $0$ or $\frac12$,
depending on whether $N$ is even or odd. 

The two-body interaction term does not commute with the free
Hamiltonian, which necessitates the introduction of time slices in
the HS transformation. Writing the imaginary-time evolution operator
as
\be
\hat U = \left[\exp(-\Delta\beta \hat H)\right]^{N_t}
\ee
where $\beta = N_t\Delta\beta$, we can apply the Hubbard-Stratonovich
transformation to obtain:
\be
e^{-\Delta\beta \hat H}\propto\int {\cal D}\sigma_x\,{\cal
  D}\sigma_y\,\exp({\textstyle-\frac12\Delta\beta\,
  V}\sum_n(\sigma_x^{(n)2}+\sigma_y^{(n)2}))\;
\Pi_n\exp({-\Delta\beta  \,\hat h_\sigma^{(n)}})\;,
\ee
where
\be
\hat h_\sigma^{(n)} = \hat J_z +\sqrt2V(\hat
J_x\sigma_x^{(n)}+{\rm i}\hat J_y\sigma_y^{(n)})\;.
\ee
With this decomposition, $\exp(-\Delta\beta\,\hat h_\sigma)$ is
accurate to order $(\Delta\beta)^2$ and $\hat U$ to order
$\Delta\beta$. Exact results are obtained in the limit
$\Delta\beta\rightarrow0$. 

To obtain averages at finite temperature, we need to take traces of
such operators over the many-body basis. 
For small $N$, the model can easily be diagonalized, which we use to
our advantage to compare Monte Carlo and Langevin
calculations of expectation values with the known exact results. We
start with the expressions for the static-path approximation
(SPA), i..e, for a single time-slice.
With the representation
\be
h_\sigma =\left(\begin{array}{cc}\frac12 & 
\frac V{\sqrt2}(\sigma_x-{\rm i}\sigma_y)\\ 
\frac V{\sqrt2}(\sigma_x+{\rm i}\sigma_y) &
    -\frac12 \end{array}\right)
\ee
we have
\be
\tra_1(e^{-\Delta\beta h_\sigma})= 2\,\cosh(\frac{\Delta\beta}2\,W)\;,
\label{oneparttrace}
\ee where
\be
W=\sqrt{1+2V^2(\sigma_x^2-\sigma_y^2)}\;, \label{arg}
\ee 
and indeed quite generally
\be
\tra_N\exp(-\Delta\beta\,h_\sigma) =
\left[2\cosh(\frac{\Delta\beta}2\,W)\right]^N\;.
\ee
Similarly, we obtain for example
\be
\la \hat J_z\ra_N = -\frac{N}{2\,W}\tanh(\frac{\Delta\beta}2)\;.
\ee
The many-time-slice expressions are straightforward extensions. For
time slice $n$, define
\be
\exp(-\Delta\beta h_\sigma^{(n)})\equiv A^{(n)} =
a_0^{(n)}+\vec a^{(n)}\cdot \vec\tau
\ee
and
\be
U_\sigma = 
\exp(-\Delta\beta h_\sigma^{(1)}),\cdots, \exp(-\Delta\beta h_\sigma^{(N_t)}) 
 \equiv  u_0 + \vec u \cdot \vec \tau \;,
\ee
where $\vec\tau$ are the usual Pauli matrices and
\be
a_0^{(n)}&=&\cosh(\frac{\Delta\beta}2 W_n)\;,\\
\vec a^{(n)} &=& \frac V W_n\,\sinh(\frac{\Delta\beta}2\,W_n)
\left(\begin{array}{c}-\sqrt2\,\sigma_x^{(n)} \\
-{\rm i}\sqrt2\,\sigma_y^{(n)}\\
-\frac1V\end{array}\right)\;.
\ee

Then
\be
\tra_N (U_\sigma) = (2u_0)^N \;,
\ee and, for example, 
\be
J_z(\sigma)\equiv \tra_N (\hat J_z U_\sigma)/\tra_N(
U_\sigma) = N \frac{u_3}{2u_0}\;.
\ee
The action can be written as
\be
S_\sigma = {\textstyle\frac{\Delta\beta\,
    V}2}\left(\sum_n^{N_t}\sigma_x^{(n)2}+\sigma_y^{(n)2}\right)
 -\log(\tra_N(U_\sigma))\;,
\ee
and observables are obtained as usual. Here, we choose to examine
\be
\la \hat J_z\ra_\beta = 
\frac{\int {\cal D}\sigma_x\,{\cal D}\sigma_y \, e^{-S_\sigma}
  J_z(\sigma)}
{\int {\cal D}\sigma_x\,{\cal D}\sigma_y \, e^{-S_\sigma}}\;.
\ee

The complex Langevin equation requires the gradients of 
$\tra(U_\sigma)$ with respect to $\sigma_x$ and $\sigma_y$ for each
time slice: 
\be
\frac{\del\tra(U_\sigma)}{\del \sigma^{(n)}_x}=\tra\left(A^{(1)}\cdots
A^{\prime(n)}_x\;\cdots A^{(N_t)}\right)\;,
\ee
where $A^{\prime(n)}_x$ is the derivative of the $n$th slice matrix
$A^{(n)}$. Thus,
\be
\dot \sigma_x^{(n)}(t) = -\frac12\left(\sigma_x^{(n)} -
  N\,\frac{\del\,\tra(U_\sigma)}{\del \sigma^{(n)}_x}/\tra(U_\sigma)
\right)\,+\eta
\ee
since 
\be
\frac\del{\del \sigma_x^{(n)}}\tra(\hat U_\sigma)^N=
N\,\tra\left(A^{(1)}\cdots A^{\prime(n)} \cdots
  A^{(N_t)}\right)\cdot\tra(U_\sigma)^{N-1}
\ee
 if we choose a basis of
multi-particle product states, and similarly for $\sigma_y$. 
In this manner, all remaining
traces are over the two-dimensional single-particle space
only. Naturally, this equation has to be separated into its real and
imaginary parts as earlier, since $\tra_N(U_\sigma)$ can become
negative if $N$ is odd. This is most easily seen examining the single
particle trace (\ref{oneparttrace}), which becomes oscillatory if the
argument $W=\sqrt{1+2V^2(\sigma_x^2-\sigma_y^2)}$ becomes complex. 

\begin{figure}[!tbp]
\center
 \epsfig{file={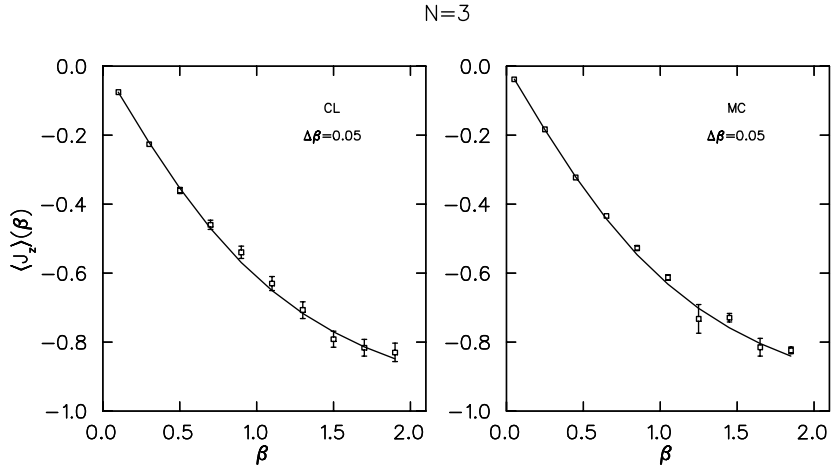}, width = .4\columnwidth, angle=90}
\caption{Monte Carlo (right panel) and complex Langevin
(left panel) average $\la J_z\ra$ in a three particle Lipkin model
as a function of inverse temperature, with $\Delta\beta=0.05$.}
\end{figure}

In Fig. 8, we show Monte Carlo (right) and complex Langevin (left)
calculations of the observable $\la J_z\ra_\beta$ as a function of the
inverse temperature $\beta$, for the first non-trivial case
$N=3$. Even though the sign is not strictly positive, the Monte Carlo
simulation is very accurate in this case. The Langevin averages
converge well also, despite the fact that none of the tricks used in
the previous examples (such as partial integration and extending the
action into the complex plane) can be used for this model. In order to
force a sign problem, we can revert to cranking as before.

\begin{figure}[!tbp]
\center
 \epsfig{file={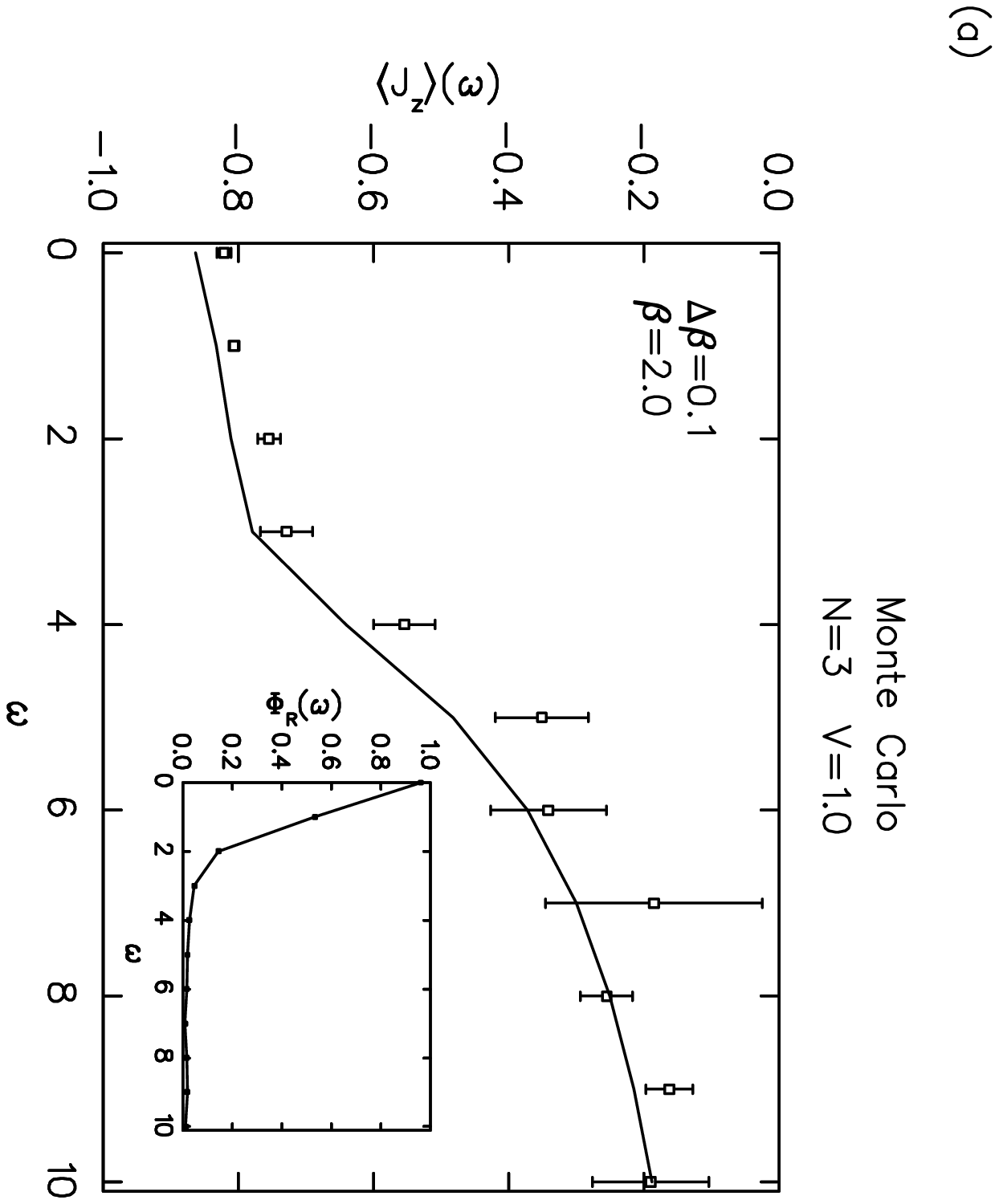}, width = .4\columnwidth, angle=90}
 \epsfig{file={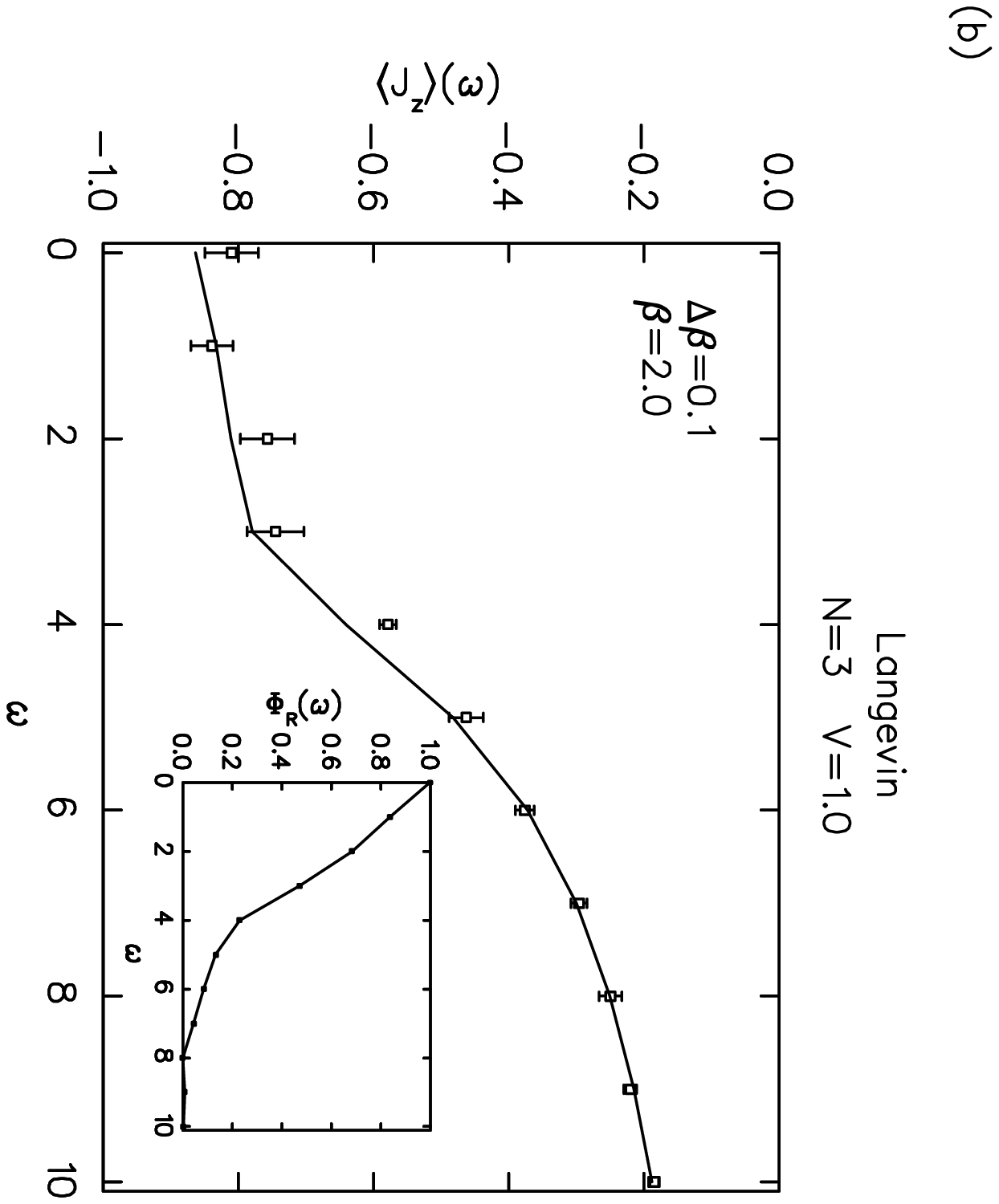}, width = .4\columnwidth, angle=90}
\caption{(a) Monte Carlo calculation of
Re($\la J_z\ra$) as in Fig.~8 (10 samples of 10,000 points), at
fixed inverse temperate $\beta=2.0$ (with $\Delta\beta=0.1$) and as a
function of cranking frequency $\omega$. The inset shows the real part
of the sign function $\Phi_R$. (b) Complex Langevin calculation of
$\la J_z\ra$ with same parameters as (a).}
\end{figure}

The cranked Hamiltonian
\be
H=H_0-\omega\, \hat J_y \label{crank}
\ee
develops a sign problem because the single particle
(and single time-slice) trace (\ref{oneparttrace}) becomes complex
as the argument (\ref{arg}) becomes
\be
W=\sqrt{1+\omega^2 + 2V^2(\sigma_x^2-\sigma_y^2) -
2\sqrt2\,{\rm i}V\omega\sigma_y}\;.
\ee
Figure 9 compares Monte Carlo and complex Langevin calculations of the
same observable as in Fig.~8, at fixed inverse temperature $\beta=2.0$
and as a function of the cranking frequency $\omega$. Because
$\tra(U_\sigma)$ becomes complex (rather than just non-positive), 
care must be given to the real and
imaginary parts of the sign function $\Phi$. Indeed, for
$\Phi=\Phi_R+{\rm i}\,\Phi_I$ and the observable $O_R(\sigma)+{\rm i}\,O_I(\sigma)$,
the Monte Carlo average is
\be
\la O\ra = \frac{\int {\cal D}\sigma \,e^{-S_\sigma}\left(\Phi_R O_R-\Phi_I
    O_I\right)+{\rm i}\int {\cal D}\sigma \,e^{-S_\sigma}\left(\Phi_I
    O_R+\Phi_R O_I\right)}{\int {\cal D}\sigma\, e^{-S_\sigma} \Phi_R +
  {\rm i}\int {\cal D}\sigma \,e^{-S_\sigma} \Phi_I} \label{expect}
\ee
Fig.~9a shows the real part of (\ref{expect}) as well as the real part
of the sign $\Phi_R$ in the inset. The sign
disappears quickly, both in the MC as well as the Langevin calculation
in Fig.~9b, but in the Langevin case the
accuracy of the average actually increases with increasing cranking
frequency. Instead, for the MC calculation, this translates into a
deteriorating signal-to-noise ratio.
Also, the complex
Langevin calculation does not necessitate the calculation of four
separate integrals such as in (\ref{expect}). However, for small
cranking frequencies $\omega$ the complex Langevin averages are
noticeably non-stationary, which results in larger error bars. 

\section{Conclusions}
The complex Langevin equation offers a new perspective on the
pervasiveness of the sign problem in fermionic quantum many-body
calculations. It is not without its own problems, however, most
notably the absence of a convergence proof of the Langevin averages.
The root of non-stationarity for some complex Langevin averages lies
in the structure of fixed points (attractors) and turning points
(repellers) in the complex plane. If both the attractors and repellers
lie on the real line, it is just a matter of time until the trajectory
hits a pole in the gradient, and the trajectory is thrown far into the
complex plane. The first two examples we have treated show how this
can be avoided by modifying the action such that the fixed points move
into the complex plane, without changing the value of the average or
the pole structure. The third example, the Lipkin model, showed that
such a procedure is not necessary if the fixed points are naturally in
the complex plane (such as is the case at finite cranking frequencies)
even though the averages may become non-stationary. In these cases,
the Langevin equation continues to deliver reliable averages even when
the Monte Carlo averages have become meaningless.

While this study certainly suggests that the sign problem can be
overcome in particular cases, it is by no means certain that the
procedure will be as successful in so-called real-life applications,
with realistic interactions. However, as the payoff is potentially
large, we believe that there is now enough evidence to try this
approach. Another area where this approach deserves to be tested is
lattice gauge calculations of matter at finite chemical potential
$\mu$, which suffer from a sign problem because the action becomes
complex. As the attractors would naturally move into the complex plane
at about $\sim {\rm i}\,\mu$, the complex Langevin approach seems particularly
natural in this case.

This work was supported in part by the National Science Foundation  
Grant Nos. PHY91-15574 and PHY94-12818, as well as a Caltech Fairchild
Fellowship to CA.


\end{document}